# Canonical-ensemble SA-CASSCF strategy for problems with more diabatic than adiabatic states: Charge-bond resonance in monomethine cyanines

*Seth Olsen**

School of Mathematics & Physics, The University of Queensland, QLD 4072, Australia

AUTHOR INFORMATION

**Corresponding Author**

*\*seth.olsen@uq.edu.au*

No abstract.

*Abstract*

This paper reviews basic results from a theory of the *a priori* classical probabilities (weights) in state state-averaged complete active space self-consistent field (SA-CASSCF) models. It addresses how the classical probabilities limit the invariance of the self-consistency condition to transformations of the complete active space configuration interaction (CAS-CI) problem. Such transformations are of interest for choosing representations of the SA-CASSCF solution that are diabatic with respect to some interaction. I achieve the known result that a SA-CASSCF can be self-consistently transformed only within degenerate subspaces of the CAS-CI ensemble density matrix. For uniformly distributed ('microcanonical') SA-CASSCF ensembles, self-consistency is invariant to any unitary CAS-CI transformation that acts locally on the ensemble support. Most SA-CASSCF applications in current literature are microcanonical. A problem with microcanonical SA-CASSCF models for problems with "more diabatic than adiabatic" states is described. The problem is that not all diabatic energies and couplings are self-consistently resolvable. A canonical-ensemble SA-CASSCF strategy is proposed to solve the problem. For canonical-ensemble SA-CASSCF, the equilibrated ensemble is a Boltzmann density matrix parametrized by its own CAS-CI Hamiltonian and a Lagrange multiplier acting as an inverse "temperature", unrelated to the physical temperature. Like the convergence criterion for microcanonical-ensemble SA-CASSCF, the equilibration condition for canonical-ensemble SA-CASSCF is invariant to transformations that act locally on the ensemble CAS-CI density matrix. The advantage of a canonical-ensemble description is that more adiabatic states can be included in the support of the ensemble without running into convergence problems. The constraint on the dimensionality of the problem is relieved by the introduction of an energy constraint. The method is illustrated with a complete active space valence-bond (CASVB) analysis of the charge/bond resonance electronic structure of a monomethine cyanine: Michler's hydrol blue. The diabatic CASVB representation is shown to vary weakly for "temperatures" corresponding to visible photon energies. Canonical-ensemble SA-CASSCF enables the resolution of energies and couplings for all covalent and ionic CASVB structures contributing to the SA-CASSCF ensemble. `The CASVB solution describes resonance of charge- and bond-localized electronic structures interacting via bridge resonance superexchange. The resonance couplings can be separated into channels associated with either covalent charge delocalization or chemical bonding interactions, with the latter significantly stronger than the former.

**1.Introduction**

State-averaged complete active space self-consistent field (SA-CASSCF) theory[1-3] is a significant tool of modern theoretical and computational photochemistry[4-5]. It is among few quantum chemical approaches that can provide a balanced description of electronic structure in systems with multiple quasi-degenerate configurations and/or strong non-adiabatic couplings[6-7].

In "typical" SA-CASSCF applications, the variational objective is the expected energy evaluated on a subspace of low-energy eigenstates. The variational state is not any one of the adiabatic states, but an ensemble of adiabatic states weighted by classical probabilities, with support on the target space. The *a priori* probabilities, as well as the number of adiabatic states in the ensemble support, are given as input parameters and fixed for a single SA-CASSCF run.

The introduction of classical statistics (state-averaging) to CASSCF was proposed as a practical tool to avoid variational collapse in models of multiple excited states with the same symmetry[2]. In reviews of CASSCF, the introduction of classical statistics is treated briefly, and the interpretation of SA-CASSCF solutions is treated as a straightforward extension of the interpretation of the solution of single-state CASSCF problems.

This paper is about a development in the theory of classical probabilities in SA-CASSCF models. The classical probabilities are important because they limit the range of complete active space configuration interaction (CAS-CI) transformations that can be used to self-consistently represent the SA-CASSCF ensemble (and its effective Hamiltonian).

The paper is organized as follows: In the next section, I briefly review the introduction of classical probabilities in SA-CASSCF models. I achieve the known result that the solution can only be self-consistently transformed within degenerate subspaces of the ensemble CAS-CI density matrix. In section 3, I review the definition of a 'microcanonical' (i.e. uniformly-weighted) SA-CASSCF. I achieve the known result that the solution can be self-consistently transformed by any unitary acting locally on the CAS-CI ensemble support. I describe a problem with microcanonical SA-CASSCF for problems with "more diabatic than adiabatic states". The problem is that not all relevant diabatic energies and couplings can be self-consistently resolved. In section 4, I propose a canonical-ensemble approach that solves this problem. Canonical-ensemble SA-CASSCF, like microcanonical-ensemble SA-CASSCF, is invariant to transformations that act locally on the ensemble support. The advantage of canonical-ensemble SA-CASSCF is that more adiabatic states can be included in the ensemble support without running into stability problems. In section 5, I illustrate this with a complete active space valence-bond (CASVB) analysis of a monomethine cyanine, that resolves all covalent and ionic energies and couplings self-consistently. I sub-divide the example as an included mini-paper. The CASVB analysis describes superexchange resonance between structures with distinguishable charge- and bond-localization, as in traditional resonance models of dyes. The resonance couplings can be decomposed into channels associated with covalent charge delocalization or covalent bond formation, with the latter significantly stronger than the former. Section 6 concludes.

**2. SA-CASSCF ensembles and their invariances**

The objective function that is optimized in a typical SA-CASSCF run (with fixed *a priori* probabilities) is the expected energy $\langle \mathbf{H} \rangle$ over an ensemble of low-energy eigenstates in the Hilbert space generated by the complete active space configuration interaction (CAS-CI) expansion:

$$(1.1) \qquad \langle \mathbf{H} \rangle = \mathrm{Tr}(\bar{\mathbf{\Gamma}} \mathbf{H}).$$

Here $\bar{\mathbf{\Gamma}}$ is the state-averaged density matrix of the SA-CASSCF ensemble and $\mathbf{H}$ is the effective Hamiltonian. For simplicity, I project out the closed portion of the wavefunctions (so the states are labeled by their active space configurations). I a basis of CAS-CI states $|\Phi\rangle$, the CAS-CI ensemble density matrix has elements

$$(1.2) \qquad \bar{\Gamma}_{AB} = \sum_{I}^{M \leq N} \langle \Phi_A | \Psi_I \rangle p_I \langle \Psi_I | \Phi_B \rangle,$$

where $|\Psi_I\rangle$ is the *I*th CAS-CI eigenvector, $|\Phi_A\rangle$ is the *A*th many-body basis state generated by the CAS-CI (at convergence of the CASSCF), and the $p_I$ are classical probabilities.

The *support* of $\bar{\mathbf{\Gamma}}$ is spanned by the adiabatic states with non-vanishing probability $p_I \neq 0$. The probabilities are given as *a priori* input parameters for a typical SA-CASSCF calculation.

The limit on the sum in (1.2) highlights that the support of $\bar{\mathbf{\Gamma}}$, which has dimension *M*, may be restricted to a subspace of the full many-particle Hilbert space generated by the CAS-CI, which has dimension $N \geq M$. It is $M$, not $N$, that restricts the range of achievable self-consistent CAS-CI representations of the effective Hamiltonian.

A CASSCF eigenstate is parameterized as[8]

$$|\Psi_I\rangle = \exp(\sum_{j \neq i} k_{ij}\mathbf{x}_{ji})\exp(\sum_{J \neq I} K_{IJ}\mathbf{X}_{JI})|\Psi_{guess}\rangle , \tag{1.3}$$

where the $\mathbf{x}_{ij}$ are single-particle shift ($i \neq j$) operators, the $\mathbf{X}_{IJ}$ are many-particle shift operators (acting on the Hilbert space generated by the CAS-CI), $k_{ij}$ and $K_{IJ}$ are variational parameters. The single-particle operators can be orbital operators (U(N) generators) or spin-orbital operators (U(2N) generators), depending on whether a configuration state function (CSF) or Slater determinant basis is chosen to construct the CAS-CI Hilbert space. For specificity, I assume a CAS-CI basis of singlet CSFs. In SA-CASSCF, there are two conditions fulfilled at convergence at the one- and many-particle levels. The single-particle convergence criterion is

$$\frac{\partial\langle\mathbf{H}\rangle}{\partial k_{i \neq j}} = \mathrm{Tr}\left(\bar{\Gamma}\left[\mathbf{H}, \mathbf{x}_{j \neq i}\right]\right) = 0 . \tag{1.4}$$

Equation (1.4) is also known as the generalized Brillouin condition[9]. As a consequence, any CASSCF state is invariant to orbital transformations that act locally on the active space orbitals.[10] This symmetry has been used to present single-state CASSCF solutions in a complete active space valence-bond (CASVB) form for ground and excited states with different symmetry.[11]

The invariances in the CAS-CI problem, that I focus on here, are distinct from the orbital transformation symmetry described above.[10]

The many-particle CAS-CI convergence criteria for he ensemble is written analogously as

$$\frac{\partial\langle\mathbf{H}\rangle}{\partial K_{I \neq J}} = \mathrm{Tr}\left(\bar{\Gamma}\left[\mathbf{H}, \mathbf{X}_{J \neq I}\right]\right) = 0 . \tag{1.5}$$

The convergence criteria assert the ensemble $\bar{\Gamma}$ is stationary under the action of the effective Hamiltonian.[9, 12] In the adiabatic representation, the diagonal elements of $\bar{\Gamma}$ are a probability distribution over the adiabatic states. In a diabatic basis, the diagonal elements will be a probability distribution with greater Shannon entropy than the adiabatic distribution[13].

When the objective function (1.1) is invariant to a class of transformations of the Hamiltonian, these transformations can be used to self-consistently represent the effective Hamiltonian in a diabatic basis. Diabatic state representations[7, 14-16] can be extremely useful for the development of models of non-adiabatic processes[17-19], solvent interactions[15, 20-21], or to understand and communicate the results of calculation in chemically intuitive language[22].

The term "diabatic" is used more often than it is defined, so I offer a brief review. I use concepts of states that are diabatic or adiabatic with respect to an interaction, after Delos' discussion of electronic transitions in slow atomic collisions.[14] If no interaction is specified, then the interaction is the effective Hamiltonian itself. The states are adiabatic if they diagonalize the interaction and are diabatic if they do not. In this way, the concept of states that are "diabatic with respect to nuclear displacements" can also be generalized to concepts of states that are "diabatic with respect to environmental interactions".[15, 21] For the case where the interaction is the nuclear displacements, the quantum mechanical nature of the nuclear motion requires no electronic representation can generally be found that is globally adiabatic with respect to the nuclear displacements.[23] Such a representation has been called a "strictly diabatic" representation in the chemical physics literature.[23] For interactions that are well-described by a classical parameterization, such as for states diabatic with respect to a (slow) solvation interaction[21], these issues may be less important. I do not use the prefix *quasi-* or other conditionals to refer to diabatic states that are not adiabatic with respect to nuclear displacements, such as has been done elsewhere, the phrase "diabatic" being reserved for representations that are adiabatic with respect to nuclear displacements.[16, 24-25]

In typical SA-CASSCF, where the variational objective is the state-averaged energy (1.1), self-consistency is only preserved within degenerate subspaces of the CAS-CI ensemble density matrix. In the general case, SA-CASSCF solutions are not invariant to transformations of the CAS-CI Hilbert space. This limits the ability to represent the SA-CASSCF solution in a diabatic basis for general SA-CASSCF ensembles with fixed *a priori* probabilities.

### 3. Invariance of microcanonical-ensemble SA-CASSCF

It is typically the case in SA-CASSCF applications available in the literature that the classical probabilities describe a uniform uniform) ensemble over a low-energy subspace of the CAS-CI Hilbert space. This ensemble corresponds to the microcanonical ensemble of statistical thermodynamics.[26]

In a microcanonical SA-CASSCF, the density operator is proportional to the identity on its support:

$$(1.6) \quad \bar{\Gamma} = \bar{\Gamma}_{even} \equiv \frac{1}{M}\mathbf{I} = e^{-(\ln M)\mathbf{I}}.$$

The last expression on the far right of (1.6) is included to highlight that a uniform distribution has the form of a maximum-entropy state constrained by a projection on the Hilbert space[27-28]. The relevant invariant Hilbert space is defined by the resolution of the identity operator in (1.6), which has dimension $M \leq N$. When the support is restricted to a subspace of the complete CAS-CI Hilbert space, then $\mathbf{I}$ is defined as a projection onto the support (and not the identity on the complete CAS-CI space). That is, I have $\mathbf{I} = \sum_{I}^{M} \mathbf{X}_{II}$, where $\mathbf{X}_{II}$ is the population operator of the $I$th adiabatic state in the support and $p_I = \text{Tr}(\bar{\mathbf{\Gamma}} \mathbf{X}_{II})$.

Maximum entropy states are "least-biased" states, in the sense that they contain no information that is not present in the constraints that specify them[13, 29]. If the number of relevant adiabatic states in the problem can be taken *a priori*, then microcanonical-ensemble weighting is a physically defensible strategy because it requires the least additional assumption.

The microcanonical SA-CASSCF objective function is the average energy:

$$\langle \mathbf{H} \rangle_{even} = \text{Tr}\left(e^{-(\ln M)\mathbf{I}} \mathbf{H}\right) = \frac{1}{M} \text{Tr}(\mathbf{IHI}) \ . \tag{1.7}$$

I have written the rightmost expression of (1.7) to highlight that only states in the support of $\bar{\mathbf{\Gamma}}$ are counted.

If the number of target states $M$ is less than the dimension $N$ of the CAS-CI Hilbert space then (1.7) is only invariant with respect to unitaries acting locally on the support of $\bar{\mathbf{\Gamma}}$. An immediate consequence is that the *total* number of self-consistent diabatic states can never exceed the *total* number of adiabatic states in the support of $\bar{\mathbf{\Gamma}}$.

*A problem with "more diabatic than adiabatic states"*

The requirement that diabatic transformations act locally on the support of $\bar{\mathbf{\Gamma}}$ is practically restrictive for microcanonical SA-CASSCF because, in practice, microcanonical SA-CASSCF calculations behave poorly when the number of adiabatic states in the ensemble exceeds very few. A problem is that the density of locally "ionic" states, with large local occupation fluctuations, rises quickly up the spectrum. The orbital structures appropriate for describing covalent and ionic states are different[30-32], and it is difficult to accommodate both within a single self-consistent field. This problem can sometimes be fixed by expanding the active space to allow more flexibility, but the price is severe in terms of computational tractability and human intelligibility.

The restriction to transformations local on the support presents a problem where there are more diabatic states than adiabatic states that are relevant to the problem. The problem is that not all relevant diabatic states and energies may be self-consistently resolvable. This would be expected wherever multiple weakly coupled diabatic states contribute to a few adiabatic states of interest. Cases have recently been highlighted for diabatic states with respect to displacements along a reaction coordinate[33], and states diabatic with respect to a substituent parameter[34]. In such cases, it may be practically impossible to obtain a well-behaved microcanonical SA-CASSCF ensemble with enough support to resolve all the relevant diabatic states. If $\bar{\Gamma}$ does not have support on the complete CAS-CI Hilbert space (i.e. if $M < N$), one can only resolve diabatic states to within a least-squares block-diagonalization transformation consistent with the structure of the support of $\bar{\Gamma}$ [24, 35]. This approach has been used in previous work to generate three-state diabatic Hamiltonians for symmetric cationic diarylmethanes[34] and green fluorescent protein chromophores[36].

A result from quantum information theory says that the Shannon entropy is always minimized in the adiabatic representation[13]. The probability distribution over adiabatic states majorizes[37] the distribution over any diabatic representation. This implies that the phenomenon of having more relevant diabatic than adiabatic states is more likely to be the rule than the exception.

Complete active space valence-bond decompositions allow the representation of CASSCF states in terms of valence-bond structures defined over localized orbitals.[38] In these models, chemical bonding is described using an interaction between covalent and ionic diabatic states. In valence-bond treatments of the hydrogen molecule using orthogonal valence-bond models, covalent and ionic states were found to represent optimal diabatic states for the system.[39] The representation of CASSCF states as superpositions of CASVB structures has led to significant insights. However, one can never resolve more diabatic states than adiabatic states by a unitary transformation of the ensemble support. CASVB descriptions of excited state electronic structures with different symmetry from the ground state have been reported.[11, 40] For excited states with arbitrary symmetry, one must typically resort to state averaging, and the discussion in this paper becomes relevant.

Ionic CASVB structures are typically higher in energy than covalent structures, due to the strong local charge fluctuations.[41] CASSCF models suffer systematic errors in the calculation of the relative energies of covalent and ionic CASVB states, which may be corrected by a subsequent dynamical correlation treatment.[41] The transformation invariance of subsequent treatments raises its own issues, which I will not consider further in this paper.[42]

**4. Invariance of canonical-ensemble SA-CASSCF models**

In this section, I propose that canonical-ensemble SA-CASSCF models offer a way around the "more diabatic than adiabatic states" problem encountered with microcanonical SA-CASSCF. Canonical-ensemble SA-CASSCF has a similar invariance to the microcanonical case, but has the advantage that the support of the ensemble can be enlarged without running into the convergence problems typical for microcanonical-ensemble SA-CASSCF with more than a few states. I will show that the benefits derive from relieving the constraint on the identity in Equation (1.6) in favor of a constraint on the Hamiltonian.

For a canonical-ensemble SA-CASSCF, the ensemble density matrix at equilibrium is a Boltzmann density matrix with the effective Hamiltonian as sufficient statistic. The distribution is parameterized by a Lagrange multiplier, which plays a role analogous to an electronic "temperature". Self-consistency implies the classical probabilities are allowed to vary; the *a priori* and *a posteriori* probabilities generally differ. At equilibrium, the ensemble is represented as

(1.8) $$\bar{\Gamma} = \bar{\Gamma}_\beta \equiv e^{-\beta(f\mathbf{I}+\mathbf{H})},$$

where $f$ is the free energy, and $\beta$ is the Lagrange multiplier associated with a constraint on the energy. The inverse $\frac{1}{\beta}$ is an energy scale, which plays the role of an effective "temperature" for the ensemble. I use quotes to prevent confusion with the physical temperature, *which it is not*. The introduction of an effective "temperature" in a spectroscopic model is in the spirit of Jaynes' predictive statistical mechanics[13, 29]. Canonical thermostatistical states (Boltzmann states) are the least biased stationary states consistent with an energy scale.

There is also the question of variation of properties with the parameter; models with a small variation of structure over a wide range of the "temperature" will provide the most credulity in applications. For the example below, this is indeed the case for "temperatures" at visible photon energy.

Taking the logarithm of Equation (1.8) and rearranging leads to a thermodynamic identity, which defines the invariant Hilbert space.

(1.9) $$\mathbf{I} = \frac{-1}{f}\left(\mathbf{H} - \frac{1}{\beta}\mathbf{S}\right),$$

where $\mathbf{S}$ is a CAS-CI *surprisal*[43-45] matrix

(1.10) $$\mathbf{S} \equiv -\ln \bar{\Gamma}_\beta.$$

The surprisal is the "operator" whose expectation value is the von Neumann entropy

$$(1.11) \quad \langle S \rangle = \text{Tr}\left(\bar{\Gamma}(-\ln\bar{\Gamma})\right)$$

In canonical-ensemble SA-CASSCF, the relevant variational objective is the free energy[46]

$$(1.12) \quad f = \text{Tr}\left(\bar{\Gamma}_\beta\left(\mathbf{H} + \frac{1}{\beta}\mathbf{S}\right)\right).$$

*The free energy is a scalar at equilibrium.* This implies invariance to local unitary transformations of the support of the ensemble. At equilibrium, the matrix representation of equation (1.9) is satisfied element-by-element on any unitarily equivalent representation of the support of $\bar{\Gamma}$. The identity in (1.9) is restricted to the support of $\bar{\Gamma}$, because the logarithm in (1.10) cannot be defined outside the support.

Equations (1.6) – (1.12) show that *both* microcanonical- and canonical-ensemble SA-CASSCF are invariant to unitary CAS-CI transformations that act locally on the support of the ensemble.

Both the microcanonical and canonical ensembles are thermostatistical ensembles, meaning they maximize the von Neumann entropy under constraint by the effective Hamiltonian or an operator that commutes with it. Using the surprisal, we can write the von Neumann entropy maximization condition as

$$(1.13) \quad \frac{\partial \langle S \rangle}{\partial (\ln p_I)} = \frac{\partial \langle S \rangle}{\partial K_{II}} = \text{Tr}\left(\bar{\Gamma}[\mathbf{S}, \mathbf{X}_{II}]\right) = 0 .$$

For the microcanonical case, this simplifies to

$$(1.14) \quad \frac{\partial \langle S \rangle}{\partial (\ln p_I)} = (\ln M)\text{Tr}\left(\bar{\Gamma}[\mathbf{I}, \mathbf{X}_{II}]\right) = 0 .$$

This result is trivially fulfilled at convergence, since we can write the identity on the support as

$$\mathbf{I} = \sum_I^M \mathbf{X}_{II} .$$

In the canonical case, the analogous expression is

$$\text{(1.15)} \qquad \frac{\partial \langle \mathbf{S} \rangle}{\partial (\ln p_I)} = \beta \operatorname{Tr}\left( \bar{\mathbf{\Gamma}} [\mathbf{H}, \mathbf{X}_{II}] \right) = 0 \ ,$$

This expression has the form of an extension of the convergence criterion (1.5) to the adiabatic population operators $\mathbf{X}_{II}$.

The key advantage of using a canonical ensemble, with respect to the problem of "more diabatic than adiabatic" states, is that the support of $\bar{\Gamma}$ can be made larger without introducing stability problems that plague microcanonical SA-CASSCF. In effect, the imposition of a constraint on the energy scale (through the effective electronic "temperature") allows relaxation of the constraints on the dimensionality of the support, which is limiting for the microcanonical case. This allows greater transformation flexibility without compromising variational focus at a well-defined energy scale.

In the example application I describe below, it was possible to equilibrate the complete CAS-CI Hilbert space, so that $\bar{\Gamma}$ has support on the complete space. If $\bar{\Gamma}$ does not have support on the complete CAS-CI Hilbert space (i.e. if $M < N$), one can only resolve CSF matrix elements to within a block-diagonalization transformation consistent with the structure of the support of $\bar{\Gamma}$ [24, 35].

The use of the full CAS-CI Hilbert space was feasible here because the active space is *small*. In applications with a much larger Hilbert space, it might be more reasonable to consider ensembles constrained to a subspace of the complete Hilbert space. Equation (1.9) still applies, but with the identity understood as a projection onto the support of $\bar{\Gamma}$. In such a situation, the available diabatic transforms would be restricted to local transformations on the support of $\bar{\Gamma}$, and the resolution would be limited to the support of the ensemble·

Moving from the microcanonical to the canonical ensemble allows us to relax the constraint on the dimensionality of the Hilbert space at the cost of introducing an energy parameter. The choice of energy parameter should be motivated by the problem to be examined. For the example below, the system is a coloured dye, and the relevant energy scale is in the visible spectrum. For photochemical problems, it is reasonable to set the thermal energy scale near the electronic gap of the molecule, or a closely related molecule. This is in the spirit of heat engine models of lasers with gain, where the temperatures of the reservoirs is determined by the energy of the relevant transitions in the lasing material[47].

For the example in this paper, I will exploit the invariance of canonical-ensemble SA-CASSCF to obtain a CASVB decomposition of the electronic structure of a dye at its ground state minimum. I do not consider specifically the problem of nuclear displacements. For this case, is is known that conical intersections between Born-Oppenheimer electronic states occur at low energy in molecules[48], and this means that no electronic representation can be found that is globally adiabatic with respect to the nuclear displacements (i.e. a "strictly diabatic" representation[23]).

Canonical SA-CASSCF models will still suffer the same problems as other SA-CASSCF models if the molecular structure changes so as to qualitatively change the character of the relevant low-energy states at some "temperature". For SA-CASSCF ensembles with support on the full CAS-CI space, these effects should be carried entirely by the orbitals, as there will be no derivative couplings arising from truncation of the CAS-CI support. In this case that the orbital structure breaks down, one may be forced to consider enlarging the active space, and the support in the enlarged space, to incorporate the relevant states in a global description. This point is well illustrated by a recent study of vibronic coupling models for ethylene, which indicate a 17-dimensional model Hilbert space to describe the low-energy electronic structure over large, but relevant, nuclear displacements.[49-50]

**5. An Example Application**

*5a. Charge/bond-resonance in Michler's hydrol blue (HB)*

To demonstrate the utility of canonical-ensemble SA-CASSCF strategies, I illustrate with a diabatic CASVB analysis of charge- and bond- resonance electronic structure in the "typical" monomethine cyanine (diarylmethane) dye Michler's hydrol blue (c.f. Scheme 1).

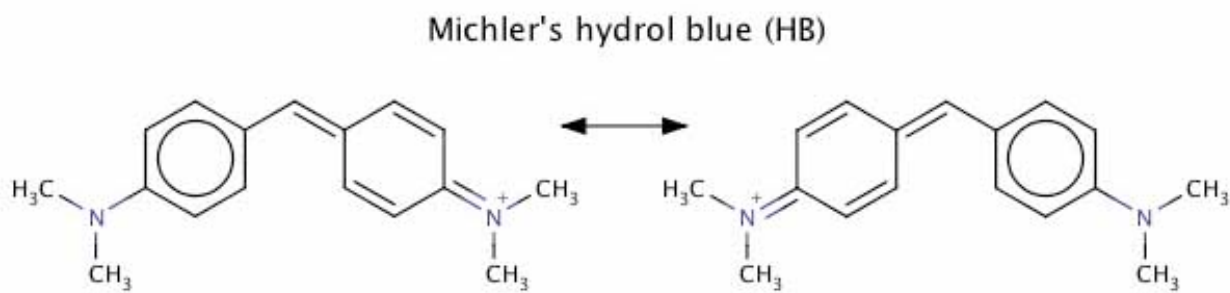

**Scheme 1**

Michler's hydrol blue (HB) has a cyanine-type chromophore.[51] This implies that its strong electronic absorbance band in the visible spectrum can be qualitatively understood using a Hilbert space model based on degenerate diabatic states with opposing charge and bond-order localization (c.f. Scheme 1)[52]. Though HB is a cation, the excess charge is electronic, because the dimethylaniline rings are bonded to the bridge in their oxidized state (and aromatic in their reduced state).

The single-particle transfer matrix element between frontier orbitals on different rings will be small in HB[53]. It is too small to explain the observed strong visible absorption near 2.0 eV[54]. Correct description of the coupling requires configuration interaction between the "canonical pair" of resonating charge/bond structures (c.f. Scheme 1) and higher-energy "intermediate" structures with different configurations of the charge and pairing degrees of freedom[55-56]. The essential physics was described by Moffitt, who invoked the formamidinum cation as an explicit example[57]. The π orbital system of the formamidinum cation has four π electrons distributed amongst three *p* orbitals. This is the *minimal complete quantum system* that can describe coherent transport of an excess electronic charge and a covalent bond-pair.

In Moffit's model, interaction between the intermediate states and the canonical resonating structures (c.f. Scheme 1) gives rise (at second order) to a strong effective coupling consistent with a visible absorbance[57]. This is a good demonstration of a canonical ensemble SA-CASSCF strategy because the relevant diabatic states (contributing to the first adiabatic gap) are high in energy and include "ionic" structures (c.f. Figure 1), but the number of relevant adiabatic states is just two.

## 5b. SA-CASSCF calculations and CASVB decomposition

I obtained solutions to the four-electron, three-orbital SA-CASSCF problem using a cc-pvdz basis set with the Molpro software[58], at the ground state minimum geometry given by Møller-Plesset second-order perturbation (MP2) theory[59] and the same basis set. The ground state geometry has been discussed previously[34, 60-62]. The weights were determined self-consistently as a canonical distribution over the six-dimensional Hilbert space of configurations of active electrons. Equilibration of the SA-CASSCF weights was implemented using an iterative loop in Molpro's internal scripting language, wherein the *a priori* weights of each SA-CASSCF calculation were calculated using state energies from the previous calculation, and the process was iterated to convergence. Calculations were performed for electronic "temperatures" (over)covering the visible spectrum (1.5-3.5 eV). This parameter region is reasonable, since we are modeling of the color of a dye.

To generate diabatic representations for the effective Hamiltonian, I chose a basis of CSFs over localized active orbitals. The orbitals were obtained by unitary transformation of the active orbitals at each "temperature" to minimize the least-squares distance from a reference set of localized active orbitals, which were the same for all "temperatures". The reference orbitals were obtained as the Boys localized active orbitals from an evenly-weighted two state SA-CASSCF calculation with the same active space structure. The two-state solution for HB has been described in previous work[60-61]. The converged active spaces were visually indistinguishable from Fig. 1 at all "temperatures". The valence structure of the solution (c.f. Fig. 1) was found to be robust for electronic "temperatures" throughout the visible.

The structure of the active space I use for Michler's hydrol blue shown in Figure 1 is analogous to the $\pi$ orbital system of formadinium cation, which was the model used by Moffitt in his configuration-interaction study of symmetric cyanines[57]. Since the diabatic basis is defined by the orbitals (via their definition as CSFs over the orbitals) the analogy extends to the many-body states as well. These are highlighted at the bottom of Figure 1. The canonical-ensemble SA-CASSCF calculations were equilibrated over the complete six-dimensional Hilbert space of configurations of active electrons in active orbitals, so I am able to resolve energies and couplings for all local-orbital CSFs spanning the space.

It is well known that SA-CASSCF gives electronic gaps that are too large by up to 2-3eV[1]. This is because of electron correlations that are not described in the CAS expansion[41]. A common way to treat these correlations is by applying a multi-reference perturbation theory correction to the SA-CASSCF reference. I have previously shown that, when such a strategy is applied to the CASSCF solutions like Fig. 1 for HB, that the excitation energy reproduces the experimental electronic gap to within reasonable accuracy. I do not report perturbation theory results here because I wish to focus on the transformation properties of the underlying SA-CASSCF ensemble. Invariance in subsequent perturbation treatments is a related, though separate, issue[42].

The molecular structure used, as well as SA-CASSCF state energies and state-averaged natural orbitals and occupation numbers for calculations at different temperatures are available in a supplement.

## Structure of canonical-ensemble CASSCF solutions for Michler's hydrol blue

### Localized active space orbital structure

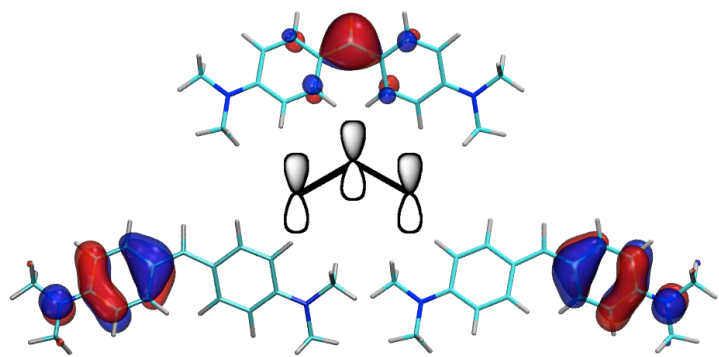

### Configuration state labels

*"Ionic" configurations*

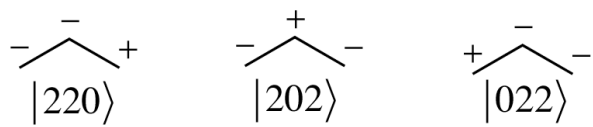

*"Covalent" configurations*

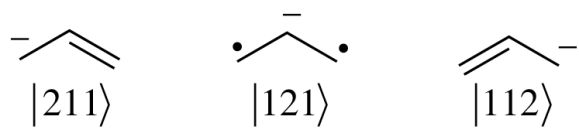

**Figure 1.** Structure of canonical-ensemble SA-CASSCF calculations for Michler's hydrol blue. Calculations used an active space of four electrons in three orbitals, which is the *minimal* complete quantum many-body model that describes coherent transport of an excess electronic charge and a bond-pair. *(Top)* Boys-localized active space orbitals for an evenly-weighted 2-state averaged CAS(4,3) calculation with a cc-pvdz basis set. These orbitals were used as a reference to determine diabatic orbitals in the canonical-ensemble calculations. Canonical-ensemble calculations used the same active space, basis set and geometry. Localized orbitals for canonical-ensemble calcluations were obtained by unitary transformation to minimize the least-squares distance from these. The resulting localized orbitals were visually indistinguishable from these for all electronic "temperatures" studied. The orbitals are localized on the rings and on the methine bridge. *(Bottom)* CSF labels for many-body basis states over the localized active orbitals. These states span the Hilbert space generated by the CAS-CI expansion. Four active electrons in three orbitals generate six singlet configurations. The localization of the orbitals allows an intuitive valence-bond labeling scheme, as shown. The visual labels are used in Figures 2 and 3. The bra-ket occupation number labels are used in the text to refer to the same states.

*5c. Results of example application*

The diabatic energies yielded by the canonical ensemble SA-CASSCF calculations on HB are shown in Figure 2, for electronic "temperatures" over-spanning the visible spectrum, and Figure 3 shows couplings extracted rom the same calculations. Only coupling elements with magnitudes greater than 1.0 eV are shown in Figure 3. All other couplings not shown were smaller than ±0.25eV in magnitude. The resolution of the individual CSF energies and couplings is only possible because the complete Hilbert space could be included in the ensemble support. This is easily done with a canonical weighting scheme, but very difficult with microcanonical (uniform) weighting. For this molecule and active space, the solution I follow does not persist if more than three states are included in a microcanonical ensemble[34]. This is insufficient support to self-consistently resolve the energies and couplings of all the six CSFs in the Hilbert space generated by the CAS-CI (cf. Fig. 1).

The key physico-chemical result that can be inferred from Figure 2 is that there are two chemically distinct "channels" that mediate the charge/bond-resonance. The channels are associated with coupling through two CSFs with charge symmetrically distributed about the bridge ("bridge states"). In Michler's hydrol blue, the bridge states are $|121\rangle$ and $|202\rangle$ (c.f. Figure 1). Both bridge states couple strongly to the resonating polarized covalent CSFs $|211\rangle$ and $|112\rangle$ (c.f. Fig. 1). These interactions will produce strong effective superexchange coupling between the low-energy resonating pair at second order via the products $\langle 211|H|121\rangle\langle 121|H|112\rangle$ and $\langle 211|H|202\rangle\langle 202|H|112\rangle$, leading to a charge/bond-resonance excitation band in the visible. For concreteness, this could be shown by least-squares block-

diagonalization of the 6-state ensemble onto the space spanned by $|211\rangle$ and $|112\rangle$. In Moffitt's model, only configuration $|202\rangle$ was explicitly considered as a possible superexchange channel, while the configuration $|121\rangle$ was neglected on the basis of higher energy and smaller coupling[57].

The charge/bond-resonance channels are distinguishable by their chemical role. Coupling via the covalent bridge state $|121\rangle$ can by interpreted as charge/bond delocalization, whilst coupling via the ionic bridge state $|202\rangle$ is associated with chemical bond formation. The latter connection derives from a fundamental result of valence-bond theory using orthogonal atomic orbitals[39, 63].

The matrix elements associated with the coupling channels are quantitatively distinct. They are not strongly dependent on the "temperature". Charge-delocalization (covalent) channel couplings are -1.76±0.12eV for HB. Chemical bond (ionic) channel couplings are in the range -2.93±0.19eV. It is notable that the ratio is not too far from the value of $\sqrt{2}$ which would be anticipated on the basis of spin counting[64]. The spin associated with the transferred charge must be opposite the stationary spins on both sites for covalent-covalent couplings, while there are no constraints on the spin of the transferred charge for the covalent-ionic couplings[65].

The pattern of coupling elements that I observe is consistent with a nearest-neighbors-only interaction between the ring and bridge fragments. This is a result, because the SA-CASSCF model could describe more complicated couplings. Coupling between the polarized covalent states $|211\rangle$ and $|112\rangle$ and their polarized ionic counterparts $|220\rangle$, $|022\rangle$ (c.f. Fig. 1) have nearly the same magnitude as their coupling to $|202\rangle$. This suggests a dominant role for covalent bonding forces in the low-energy electronic structure of this molecule. The interaction elements $\langle 211|H|220\rangle$ and $\langle 112|H|022\rangle$ stabilize the bonds in the resonating states but do not contribute to the effective coupling between them (through second order).

The bridge states $|121\rangle$ and $|202\rangle$ are quasi-degenerate and weakly coupled at all "temperatures". This implies that, where these CSFs contribute to the same adiabatic state, the adiabatic state will be sensitive (in energy and structure) to perturbations that change the relative energies of the contributing diabatic states.

The energy dispersions of the bridge states $|121\rangle$ and $|202\rangle$ are different. The dispersion of the chemical bonding channel is larger, heralding a dominant role for bonding forces in the low-energy electronic structure. Both intermediate states contribute strongly to the $S_2$ and $S_3$ adiabatic states in Michler's hydrol blue. Neither contributes to the $S_1$ state, by symmetry, nor the $S_4$ state. The ionic bridge state $|202\rangle$ contributes significantly to the ground adiabatic state[61], consistent with a larger energy dispersion for this state, while the contribution of $|121\rangle$ to the ground state is much less significant.

*5d. Discussion of example application*

I have demonstrated the utility of canonical-ensemble SA-CASSCF for generating diabatic electronic structure models by applying it to the problem of the large effective charge/bond-resonance coupling in a typical monomethine cyanine: Michler's hydrol blue. Understanding the electronic structure of Michler's hydrol blue is important because it is an important reference state for understanding the behavior of related dyes[66-68]. It also has specific practical application as an amyloid-sensitive fluorogenic stain[54].

The principle result of our application, with respect to the physics and chemistry of Michler's hydrol blue, is that the charge/bond-resonance superexchange coupling can be decomposed into *two channels with distinguishable chemical roles*. One resonance channel operates within the manifold of covalent structures and corresponds to delocalization of the charge. This channel has the smaller energy dispersion. The second channel, with larger dispersion, is associated with covalent bonding forces. The large dispersion of this channel is indicative of the dominance of pair bonding forces in the low energy spectrum of the molecule. Coupling is only significant for matrix elements that represent nearest-neighbor particle transfers. This is a concrete result, because SA-CASSCF does not assume this, and could describe higher-order couplings if needed.

I have showed in an earlier study[34] that substitution at the bridge of Michler's hydrol blue changes [34]significantly the character of the diabatic "intermediate" state obtained from a 3-state microcanonical SA-CASSCF. The changes were consistent with a change in the character of the diabatic state between two states with different character in an expanded Hilbert space. The distinct coupling channels could not be fully resolved, because I used a microcanonical-ensemble SA-CASSCF with only three states in the support. Solutions with $M \geq 3$ could not be obtained, preventing a representation with more diabatic states. The current results validate the interpretation that was given, then, that the relative electrodonation power of the substituent was tuning the balance between two channels with different coupling strength. The ability to resolve the component states, which was not possible using microcanonical SA-CASSCF methods, shows the power of the canonical-ensemble approach for enhancing the chemical interpretability of SA-CASSCF solutions.

The possibility of physico-chemically distinct dual resonance channels in methine systems has been independently suggested by by Sissa et al. in a semiempirical essential-states Hamiltonian model for the low-energy spectra a series of polymethines[64]. This model featured dual resonance channels with identical chemical interpretation to those I have identified here for hydrol blue.

The separation of the channels allowed the authors to describe the difference between final states reached by two-photon absorption vs. one-photon excited-state absorption. Our results can be taken as independent *ab initio* support for the parametric structure suggested by Sissa et al. for more general methine systems[64].

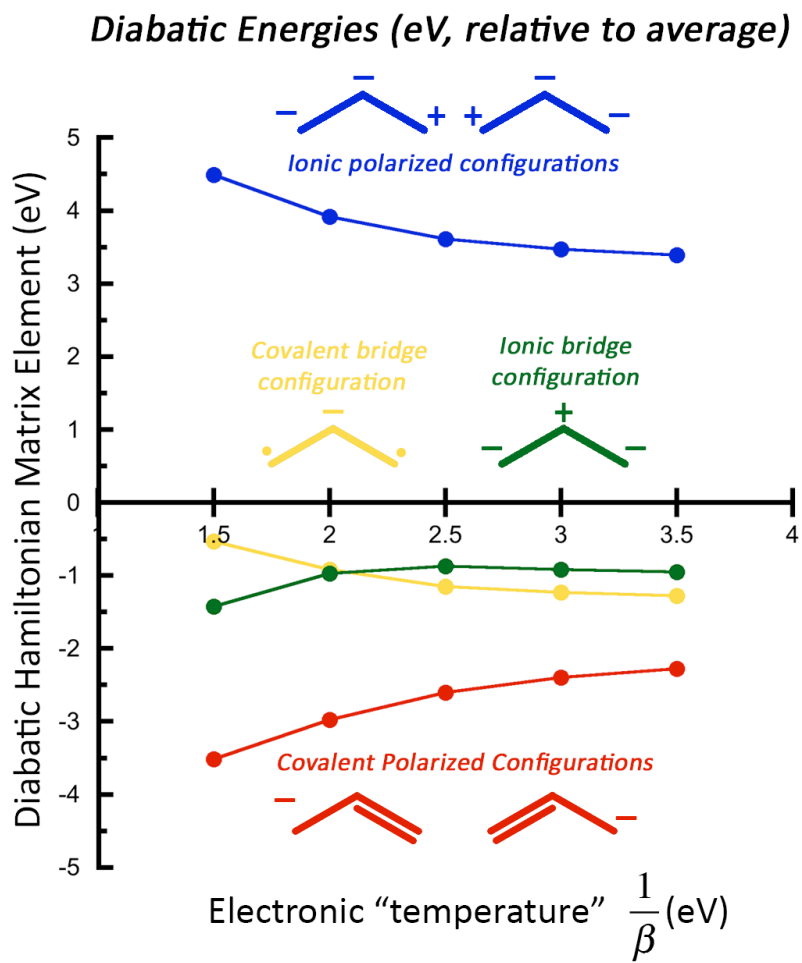

**Figure 2.** SA-CASSCF diabatic state energies (eV, relative to state-average) for Michler's hydrol blue, calculated with canonical-ensemble SA-CASSCF at electronic "temperatures" spanning the visible spectrum. The diabatic states are singlet configuration state functions (CSFs) over localized orbitals. The localized orbitals indistinguishable from Figure 1 at all "temperatures". The ensemble was equilibrated over the complete Hilbert space generated by the CAS-CI. That transformation to the diabatic basis leaves the free energy invariant and preserves self-consistency of the transformed Hamiltonian.

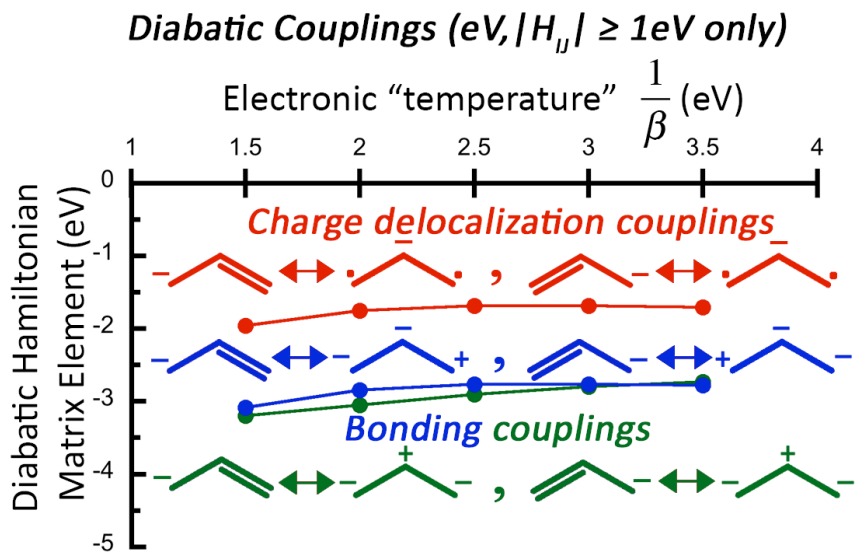

**Figure 3.** Diabatic state couplings calculated for Michler's hydrol blue using a canonical-ensemble CASSCF parametrized by electronic "temperatures" spanning the visible spectrum. Only couplings with magnitude $|\mathbf{H}_{IJ}| \geq 1.0$ eV are shown. The largest coupling not shown did not exceed 0.25 eV for any electronic "temperature" studied. The identity of the couplings is shown using valence-bond structural labels described in Figure 1.

## 6. Discussion: Canonical-ensemble SA-CASSCF for problems with more diabatic than adiabatic states

This paper reviewed some basic early results in a theory of the classical probabilities in SA-CASSCF models. The results concern the limitations set by the selection of probabilities on the range of configurational transformations that can be used to self-consistently represent the CAS-CI ensemble. I showed that both microcanonical and canonical-ensemble SA-CASSCF have the advantageous property of being invariant to unitary transformations that are local to the ensemble support. This result has been previously reported for the microcanonical case, but I am unaware of such a discussion of the canonical case. Canonical weighting of SA-CASSCF has been reported previously, but the transformation properties were not discussed.[69]

The results identify a problem with microcanonical SA-CASSCF applied to systems where there are more relevant diabatic than adiabatic states, and I suggest canonical-ensemble SA-CASSCF as a way around the problem. A theorem of quantum information theory implies problems with more diabatic than adiabatic states are the norm, not the exception.[13] Canonical-ensemble SA-CASSCF allows relaxation of the constraint on the dimensionality in favor of a constraint on the energy. It allows expansion of the ensemble support while maintaining variational focus on a well-defined energy scale, but requires specification of a "temperature". The "temperature" of the SA-CASSCF is the inverse of the Lagrange parameter conjugate to the Hamiltonian. It has energy units and is unrelated to the physical temperature.

Although SA-CASSCF has been used to generate photochemical models for decades, the theory of the classical probabilities in the method is, I believe, underdeveloped. Classical statistics were introduced to CASSCF as a practical tool to avoid variational collapse in cases where multiple states share common symmetry[2]. In well-known and comprehensive reviews of CASSCF techniques, there is little discussion of how the statistical model chosen affects the interpretation of the solution[1, 70-71]. In this paper, I have shown that the statistical model can limit the interpretation by limiting the representations that preserve self-consistency. This affects the interpretation because CAS-CI transformations are often useful to highlight the physical and chemical content of SA-CASSCF calculations.

To our knowledge, this is the first dedicated discussion of SA-CASSCF transformation invariance for statistics other than the microcanonical ensemble. The invariance of microcanonical-ensemble SA-CASSCF has been pointed out by Stålring et al, who exploited it in an analytical gradient implementation[8]. No analytic gradient implementation of canonical-ensemble SA-CASSCF exists, to my knowledge. Any such technology would seem to require extension so as to account for changes in the probabilities. Deskevich et al. implemented self-consistent SA-CASSCF weighting schemes using Boltzmann, Gaussian and logistical statistical models parameterized by an energy scale[69], but did not explicitly discuss implications for diabatic transformation of the effective Hamiltonian.

Canonical-ensemble SA-CASSCF offers a way to enlarge the ensemble support while maintaining variational focus at an energy scale. If the character of the electronic structure at the energy scale becomes ambiguous, then we expect to see breakdown similar to other SA-CASSCF models.[72-74]

At a conical intersection of adiabatic states, canonical-ensemble SA-CASSCF should give identical probabilities for degenerate states. The character of each adiabatic state changes, but so long as there are no other states interacting that are not treated at the same level, I would expect the state-averaged orbitals to change smoothly. State-averaged orbitals are often found to vary smoothly, even for non-adiabatic problems.[6]

Canonical-ensemble SA-CASSCF requires the assignment of a "temperature" that sets the energy scale of interest. This should be chosen in a way that is defensible on the physics of the problem. For the example I use here, the results vary weakly for "temperatures" corresponding to visible photon energies, which are the energy scales relevant to the question of the color of a dye. In general cases, one expects that the credulity of the results will be stronger if the solutions overlap strongly over a wide window of "temperatures". The breakdown at the limits of this window may yield further insights into specific systems, by highlighting the energy scales where particular physical models may or may not apply.

Both microcanonical- and canonical-ensemble SA-CASSCF can be considered as examples of a more general self-consistent field framework based on stationary maximum entropy states.[12] A corresponding dynamical theory also exists[75], and pursuit of this connection may be interesting in the future.

Jaynes[29] has pointed out that thermostatistics has a rigorous, mathematically equivalent and *subjective* interpretation as a system of predictive statistical inference, because thermostatistical states, being maximum-entropy states, are *least-biased predictions* consistent with thermodynamic observables.[29] This interpretation can also be extended to quantum statistics[13]. Our discussion here suggest that a subjective interpretation may also be given to SA-CASSCF models, as least-biased models of molecular electronic structure conditioned on *a priori* information in the molecular structure, the initial guess ensemble, and an energy scale.

The interpretation of SA-CASSCF solutions is usually as a weighted least-squares approximation to a subspace of the exact Hamiltonian. The point of the above paragraph is that this is not necessarily the only way to look at the problem. Another way is to look at SA-CASSCF as a method of *quantum statistical inference under constraint*. For example, instead of thinking of the ground state of our Michler's hydrol blue calculations as a least-squares approach to an exact ground state, we may think of it as the least-biased estimate of the structure of the ground state *of the charge-resonance system embedded in Michler's hydrol blue*.

**Acknowledgements**

This work was supported by Australian Research Council Discovery Project DP110101580. Computations were performed at the NCI National Facility at ANU, using time allocated under Merit Allocation Scheme grant m03. I thank R.H. McKenzie, T.J. Martínez, and R. J. Cave for comments on an early version of the manuscript.